# Predicting protein-protein interactions based on rotation of proteins in 3D-space

Samaneh Aghajanbaglo, Sobhan Moosavi, Maseud Rahgozar
School of Electrical and Computer Engineering
College of Engineering, University of Tehran
Tehran, Iran
s.janbaglo@ece.ut.ac.ir
{rahgozar, sobhan.moosavi}@ut.ac.ir

Amir Rahimi
School of Advanced Medical Sciences and Technologies
Shiraz University of Medical Science
Shiraz, Iran
rahimiamir@sums.ac.ir

***Abstract*— Protein-Protein Interactions (PPIs) perform essential roles in biological functions. Although some experimental techniques have been developed to detect PPIs, they suffer from high false positive and high false negative rates. Consequently, efforts have been devoted during recent years to develop computational approaches to predict the interactions utilizing various sources of information. Therefore, a unique category of prediction approaches has been devised which is based on the protein sequence information. However, finding an appropriate feature encoding to characterize the sequence of proteins is a major challenge in such methods. In presented work, a sequence based method is proposed to predict protein-protein interactions using N-Gram encoding approaches to describe amino acids and a Relaxed Variable Kernel Density Estimator (RVKDE) as a machine learning tool. Moreover, since proteins can rotate in 3D-space, amino acid compositions have been considered with "undirected" property which leads to reduce dimensions of the vector space. The results show that our proposed method achieves the superiority of prediction performance with improving an *F-measure* of 2.5% on Human Protein Reference Dataset (HPRD).**

## I. Introduction

Proteins are the most important biological molecules for living cells. Most proteins interact with each other in order to accomplish their biological functions. Different sets of these interactions form Protein-Protein Interaction (PPI) networks. The interactions among proteins are essential for many biological functions, such as DNA replication, protein synthesis, immunologic recognition, regulation of metabolic pathways, and progression through the cell cycle [1]. Hence, characterizing PPIs is important for understanding biological systems. Some experimental techniques such as those outlined by Shoemaker and Panchenko [2] are available for detecting protein interactions. Experimental techniques fall into two categories [3], 1) small-scale methods which determine the interactions between a small number of proteins; and 2) high-throughput methods that are able to identify a large number of interactions, including Yeast Two-Hybrid [4] and Mass Spectrometry [5]. However, these methods have some deficiencies such as high rates of false positive and false negative [6]. Furthermore, they are expensive in terms of time, cost and expertise [7]. That is why, beside the experimental techniques, computational approaches have been developed to accelerate and expand the research results with lower costs. These methods which are complementary to experimental ones, rely on various data resources [8], including network information [9-11], genomic information [12], evolutionary knowledge [13-16], structural information [17, 18], and domain information [19-21]. Since the sequence information of a protein is one of the most available kinds of information and specifies the protein characteristics, many sequence based methods have been devised to infer PPIs. Generally in such methods, sequence information of each protein is represented as a vector using a feature encoding method. Then, vectors of a protein pair are concatenated and a machine learning algorithm is employed in order to decide on the reliability of the interaction. Hence, feature encoding is an important part of sequence based approaches.

Accordingly, Shen et al. [22] proposed a sequence based method to predict protein-protein interactions using CONJOINT TRIAD for describing amino acids and a Support Vector Machine (SVM) [23] as a machine learning tool; and a following work which combined the surface concept into CONJOINT TRIAD because of a hypothesis that in protein-protein interactions, the surface amino acids are more important than the others [24]. Yu, Chou and Chang [25] proposed a method that designed a significance calculation to encode protein sequences and used Relaxed Variable Kernel Density Estimator (RVKDE) [26] to predict proteins interactions.

In this study, various N-Gram [27] encoding approaches are implemented to transform the sequences information of proteins into feature vectors. In addition, since proteins can rotate in 3D-space while preserve their characteristics, amino acid compositions are considered with two properties, "directed" and "undirected". After concatenating the vectors of a protein pair, performances of all encoding methods are evaluated using a RVKDE as a machine learning tool to predict proteins interactions.

The rest of the paper is organized as follow: in section 2, after describing the dataset, the proposed method and different feature encoding methods are introduced; then, their performances are evaluated in section 3.

## II. Materials and Methods

### A. Dataset

This study adopts 20 training and testing sets [25] which are based on the Human Protein Reference Database (HPRD) [28], Release 7. This version of HPRD contains 25,661 proteins and 38,167 interactions where some of them removed by Yu, Chou and Chang [25] as follows:

- Interactions which have more than two proteins as participator
- Interactions that contain a protein sequence with selenocysteine (U)
- Interactions which are based on in vitro experiments

Finally, 17,855 positive protein pairs with 6,429 proteins are remained, while a negative set is constructed from the 6,429 proteins. Furthermore, each protein pair in the negative set is not included in any of the 38,167 interactions. Since a negative instance is randomly sampled from the negative set, the 20 training and testing sets are generated with the same positive instances and distinct negative sets. Every training set has 16,855 interacting protein pairs and 16,855 non-interacting ones and each testing set has 1,000 interacting and 1,000 non-interacting protein pairs.

### B. The N-Gram Encoding Method

Proteins are macromolecules consisting of essentially twenty different amino acids which are arranged into a linear chain. Generally, the protein sequence determines protein's native fold and behavior. The N-Gram [27] is a well-known category of methods for encoding sequential data into feature vector which consists of frequency of each N adjacent item in the sequence. One successful implementation of such methods is CONJOINT TRIAD [22] which considers frequency of every three adjacent amino acids as a feature (a triad). With 20 different amino acids we have 20 ^ 3 = 8000 distinct features that would be difficult to handle for machine learning tools. Shen et. al. alleviated the dimensions of feature space by categorizing 20 different amino acids into seven classes according to their dipoles and volumes (TABLE I) [22]. Thus, the dimension of vectors is reduced to 7 ^ 3 = 343 features for each protein.

TABLE I. Categorizing 20 Different Amino Acids into Seven Classes According to Their Dipoles and Volumes [22].

| Class NO. | Class Members |
|---|---|
| 1 | Ala(A), Gly(G), Val(V) |
| 2 | Ile(I), Leu(L), Phe(F), Pro(P) |
| 3 | Tyr(Y), Met(M), Thr(T), Ser(S) |
| 4 | His(H), Asn(N), Gln(Q), Tpr(W) |
| 5 | Arg(R), Lys(K) |
| 6 | Asp(D), Glu(E) |
| 7 | Cys(C) |

After categorizing amino acids, each amino acid composition is scanned along the protein sequence. Therefore, in feature vector (*F*), each feature $f_i$ is the frequency of $i^{th}$ triad in the sequence. In the following, we will show this method as 3-Gram(7) in which 3 indicates the number of considered adjacent amino acids and (7) indicates the total number of different amino acid classes. As shown in Fig.1, we also implemented some other N-Gram encoding methods such as 1-Gram(20), 1-Gram(7), 2-Gram(20) and 2-Gram(7) and evaluated them in order to find the most appropriate ones. It should be noticed that since the frequency of each feature depends on the length of proteins, before concatenating the vectors of a protein pair, MinMax normalization is applied to features based on (1) where, $f_i$ means a feature and $d_i$ is a normalized feature.

$$d_i = \frac{\left(f_i - \min\{f_1, f_2, ..., f_N\}\right)}{\max\{f_1, f_2, ..., f_N\} - \min\{f_1, f_2, ..., f_N\}} \quad (1)$$

### C. Other Combined Factors

#### 1) Undirected Property

In polypeptide chains, twenty different amino acids are linked together by peptide bonds and form proteins. Since proteins can rotate in 3D-space, the direction concept in the arrangement of amino acid compositions is meaningless. Thus, in this study some encoding methods are tested considering both "directed" and "undirected" properties. For example, the ACD and DCA features are counted as the same in one test while in the other test they are counted separately. However, direction in 1-Gram is meaningless. To transform a sequence of amino acids to a feature vector with N-Gram encoding, an N-dimension matrix contains the frequencies of amino acid compositions. Regarding "undirected" property, about half of matrix elements can be eliminated. In Fig.2, as an example, in

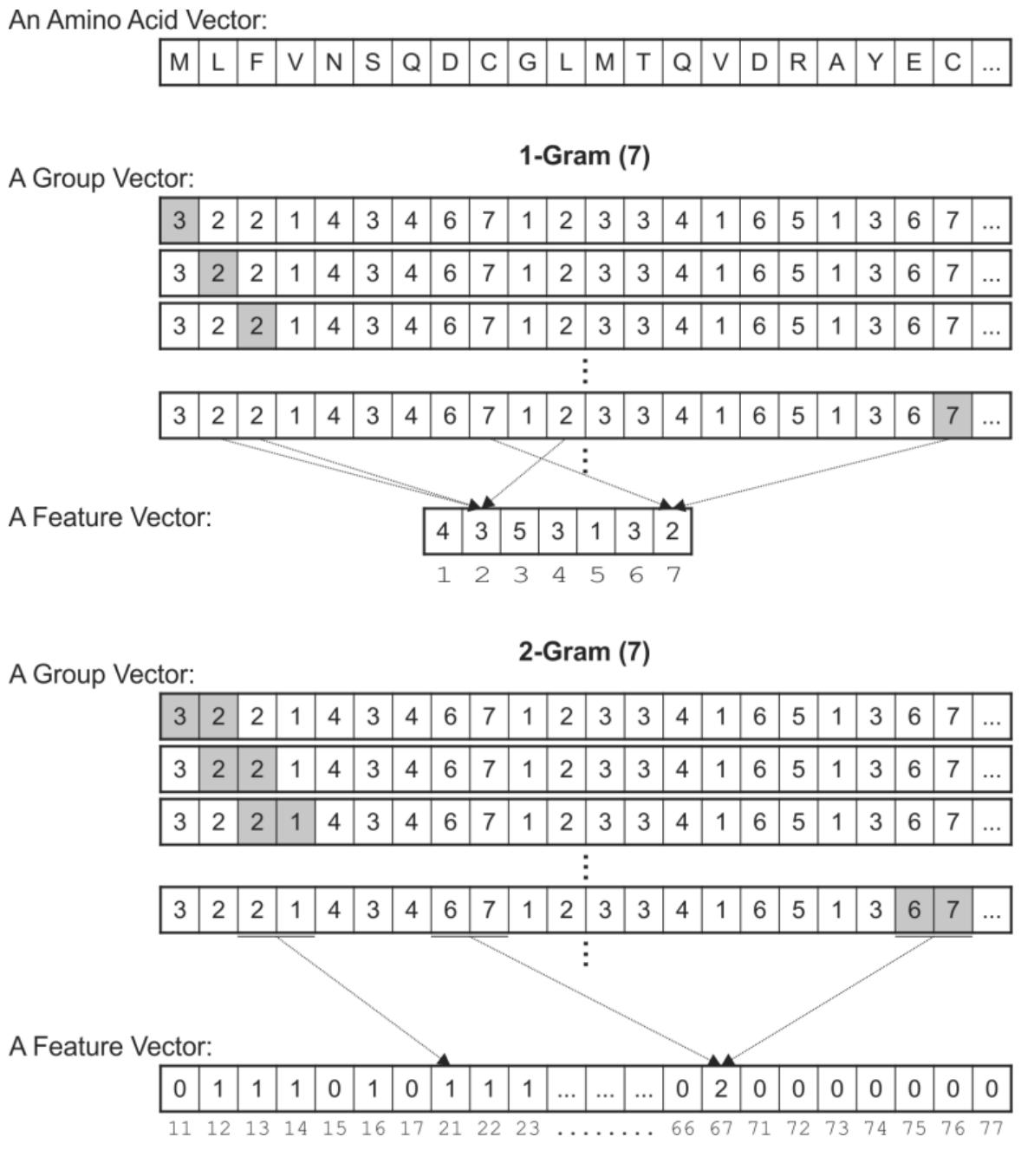

Fig. 1. An example of 1-Gram(7) and 2-Gram(7) encoding method.

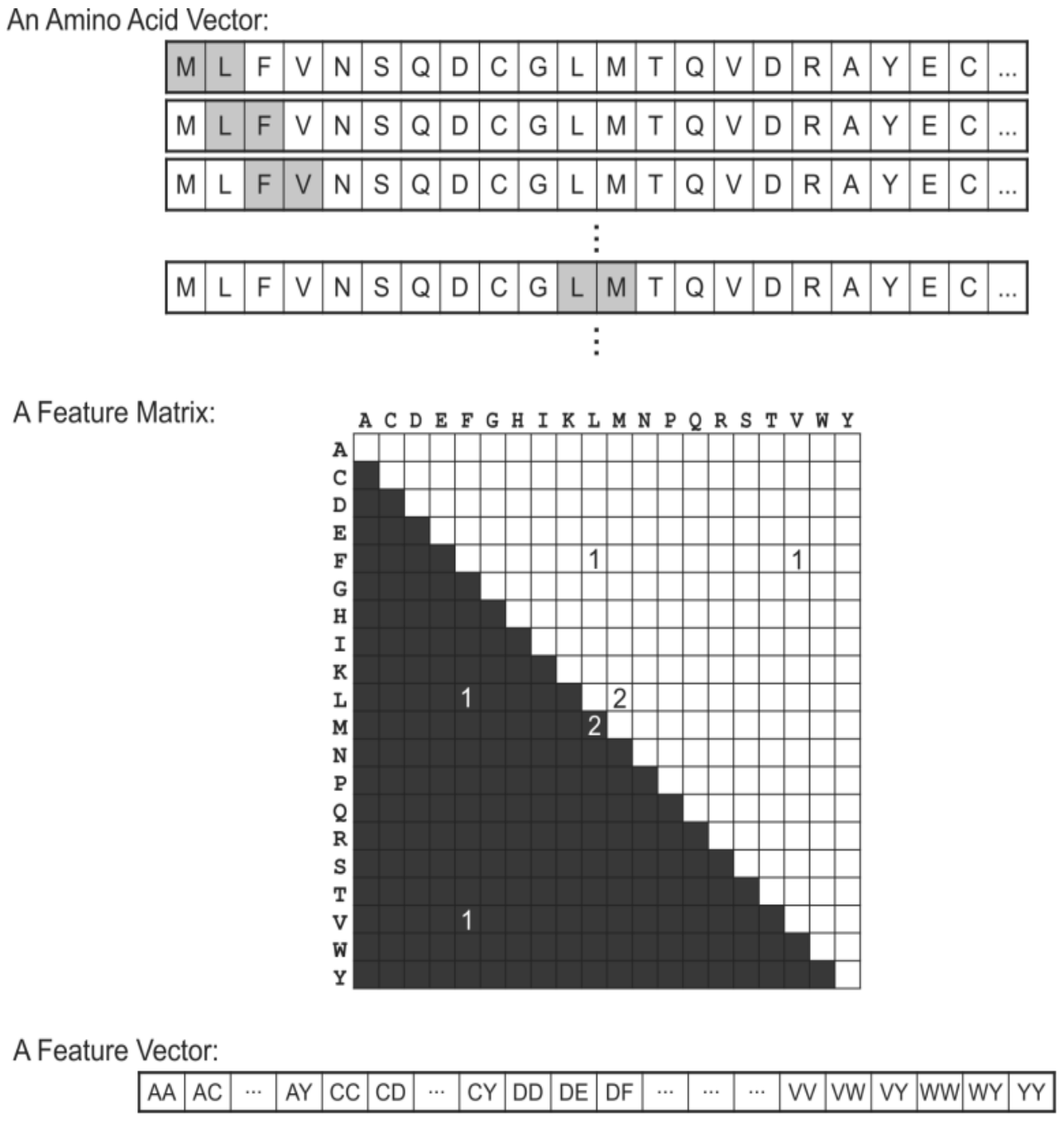

Fig. 2. An example of "undirected" property regarding 2-Gram(20) encoding.

2-Gram(20) encoding, a two-dimension matrix of 20×20 is required, and when an amino acid composition like ME is observed, the frequency of ME and EM is increased by one. All the cells below the main diagonal of the matrix are eliminated and the remaining cells are considered as the feature vector elements.

*2) Significance Vector*

Since the features' frequencies are correlated with distribution of amino acids, a probability calculation [25] was proposed to reduce the effects of amino acids distribution by transforming the feature vector into a significance vector as follows. Let $f_i$ be the $i^{th}$ feature in vector. After permutation of original sequence by 10000 times, each feature $s_i$ in significance vector is formulated by formula (2) where $x_{ij}$ is the number of the $i^{th}$ triad observed in the $j^{th}$ permuted sequence. Since, significance vector is normalized by its definition, normalization is not required.

$$s_i = \frac{1}{10000}\sum_{j=1}^{10000} t_{ij}, \mathrm{t}_{ij} = \begin{Bmatrix} 1, & x_{ij} < f_i \\ 0, & otherwise \end{Bmatrix} \quad (2)$$

## *D. Relaxed Variable Kernel Density Estimator (RVKDE)*

Relaxed Variable Kernel Density Estimator (RVKDE) [26] was proposed to estimate the probability density function of each positive or negative (interacting or non-interacting) class in the training dataset. This kernel has the ability to hand over the same level of *Accuracy* as the Support Vector Machines (SVM) [23], and requires less time to construct a classifier because of Time complexity of *O(nlogn)* order, where *n* is the number of samples in training dataset. For predicting PPIs, RVKDE creates two kernel density estimators to approximate the distributions of positive and negative protein pairs in training set. When a protein pair (ν) is sent to kernel, RVKDE predicts the class which has maximum value in likelihood function (3).

$$L_j(\nu) = \frac{|S_j|.f_j(\nu)}{\sum_h |S_h|.f_h(\nu)'} \quad (3)$$

Where, j specifies positive or negative instances. The set of class-j training samples is $S_j$ which are random and independent in the α-dimensional vector space, and $|S_j|$ is the number of class-j training samples. In addition, the kernel density estimator corresponding to class-j training instances is defined as follow:

$$\delta_i = \beta \frac{R(s_i)\sqrt{\pi}}{\sqrt[\alpha]{(\kappa+1)\Gamma(\frac{\alpha}{2}+1)}} \quad (4)$$

Where $R(S_i)$ is the maximum distance between $s_i$ and κs nearest training instances and Γ(.) is the Gamma function. In addition, $\delta_i$ is formulated based on (5). $\alpha$, $\beta$, κs and κτ are four parameters which are initializing through a five-fold cross validation.

$$\hat{f}_j(\nu) = \frac{1}{|S_j|}\sum_{s_i \in S_j}\left(\frac{1}{\sqrt{2\pi}.\delta_i}\right)^{\alpha} \exp\left(-\frac{\|\nu - s_i\|^2}{2\delta_i^2}\right) \quad (5)$$

# III. Results and Discussion

## *A. Measurements*

To evaluate the performance of feature encoding methods, some measurements are required. TABLE II lists three measurements and their formulation which are widely used in similar scopes. For PPI prediction purpose, one of the evaluation measurements is *Accuracy* that defines an overall performance of the classifier. The *F-measure* is appropriate when one class is more important than the other one; therefore, this measure is suitable for the proposed method, because the positive class attracts most attention. *Matthew's Correlation Coefficient (MCC)* is a correlation coefficient between the observed and predicted binary classifications and returns a value between −1 and +1. If the size of classes is different, the *MCC* act as a balanced measure. To realize the formulation of measurements, it is necessary to understand the following concepts:

- True Positive (TP) is the number of positive protein pairs which are correctly predicted.
- True Negative (TN) is the number of negative protein pairs which are correctly predicted.
- False Positive (FP) is the number of negative protein pairs which are incorrectly predicted as positive.
- False Negative (FN) is the number of positive protein pairs which are incorrectly predicted as negative.

TABLE II. LIST OF MEASURMENTS AND THEIR FORMULATION.

| Measurement | Formulation |
|---|---|
| Accuracy | $\frac{TP+TN}{TP+TN+FP+FN}$ |
| F-measure | $\frac{2TP}{2TP+FP+FN}$ |
| Matthew's Correlation Coefficient (MCC) | $\frac{TP\times TN-FP\times FN}{\sqrt{(TP+FP)(TP+FN)(TN+FP)(TN+FN)}}$ |

## B. *Evaluation*

One of the most available data to transform proteins to feature vectors is their sequence information, and a key part of sequence based PPIs prediction methods is how to encode the sequence information into the feature vector. Hence, In the presented work, various N-Gram encoding approaches are implemented and their performances are evaluated using a Relaxed Variable Kernel Density Estimator (RVKDE) [26] based on the Human Protein Reference Database (HPRD) [29]. For each method, the average performance of 20 test sets which have the same positive instances and distinct negative sets is reported to eliminate the dataset bias.

The results (TABLE III) show that the 2-Gram(20) encoding method achieves the best performance comparing to the other variants of N-Gram(20 or 7) with three measurements including *MCC*, *F-Measure* and *Accuracy*. It can be interpreted that in the 1-Gram methods only the amino acid composition of proteins are considered and it does not contain any information about the order of amino acids. On the other hand, in 3-gram(20) the sequence information are well encoded; but each vector has 8000 features and when vectors of a protein pair are concatenated, the feature space have 16000 dimensions which is hard to handle by the classifiers. In addition, when amino acids are categorized into seven classes to reduce the feature space dimensions, some useful information will be lost, because all members of each class are considered as identical. But, in 2-Gram(20), both the amino acid composition and the order of them are considered. Furthermore, the size of vectors is 20^2=400 features which can be easily handled by the classifier. Thus, 2-Gram(20) encoding is more appropriate rather than the other N-Gram encodings.

As mentioned previously, all encoding methods are tested with two properties of "directed" and "undirected"; this is mainly because proteins can rotate in 3D-space and direction concept in amino acid compositions is meaningless. Therefore, it is preferable to eliminate the direction of amino acid compositions which leads to reduce the dimensions of the vector space. Fig.3, 4 and 5 show the results of comparing these two properties with three measurements *MCC*, *F-Measure* and *Accuracy*.

TABLE III. RESULT OF DIFFERENT TYPES OF N-GRAM ENCODINGS.

| Features | MCC (%) | F-Measure (%) | Accuracy (%) |
|---|---|---|---|
| 1-Gram (7) | 32 | 65.1 | 66 |
| 1-Gram (20) | 51.8 | 75.8 | 75.9 |
| 2-Gram (7) | 50.3 | 75.3 | 75.2 |
| **2-Gram (20)** | **54.3** | **77.5** | **77.1** |
| 3-Gram (7) | 52.2 | 76.4 | 76.1 |

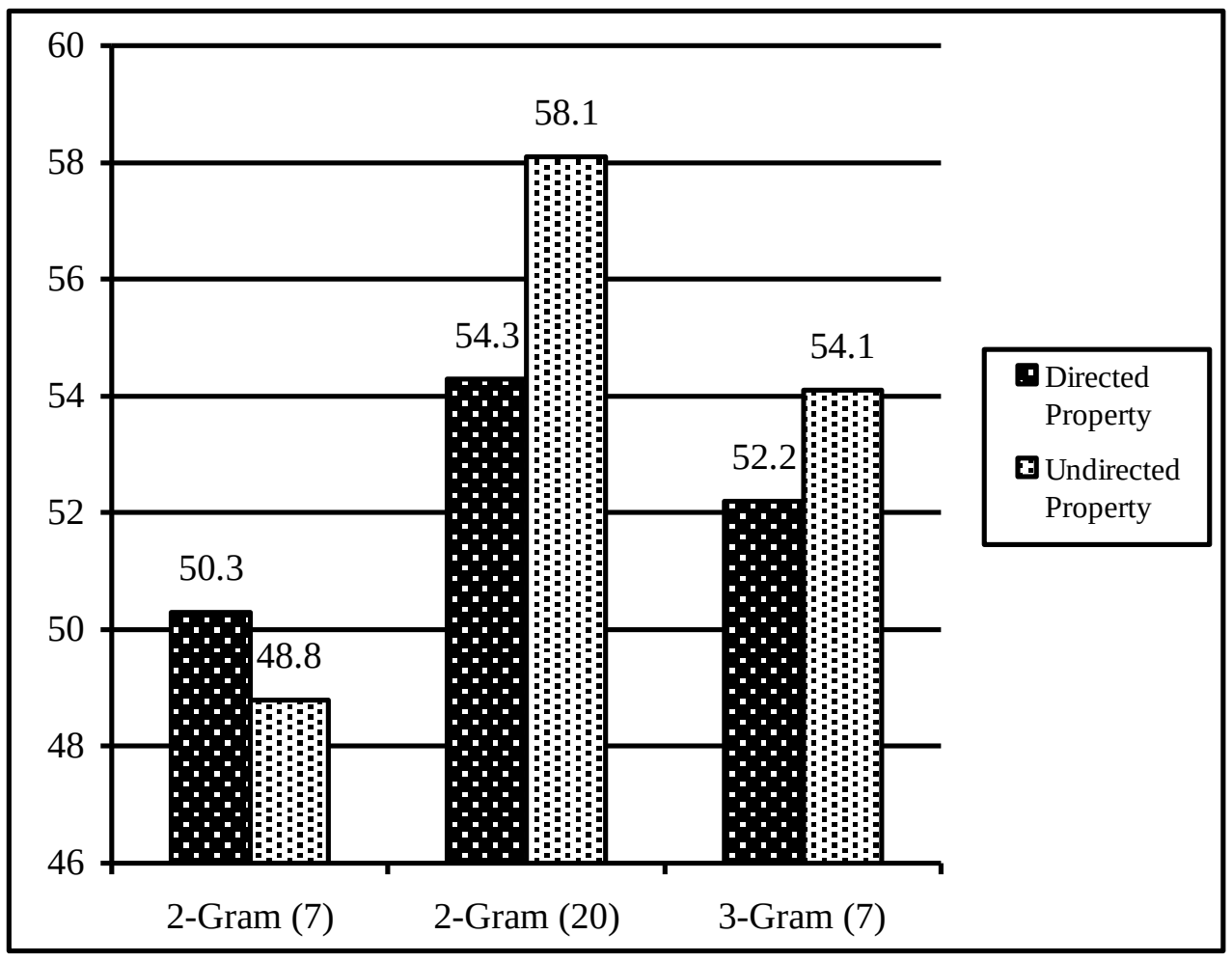

Fig. 3. Comparison of the MCC resulted by various feature encoding methods and properties.

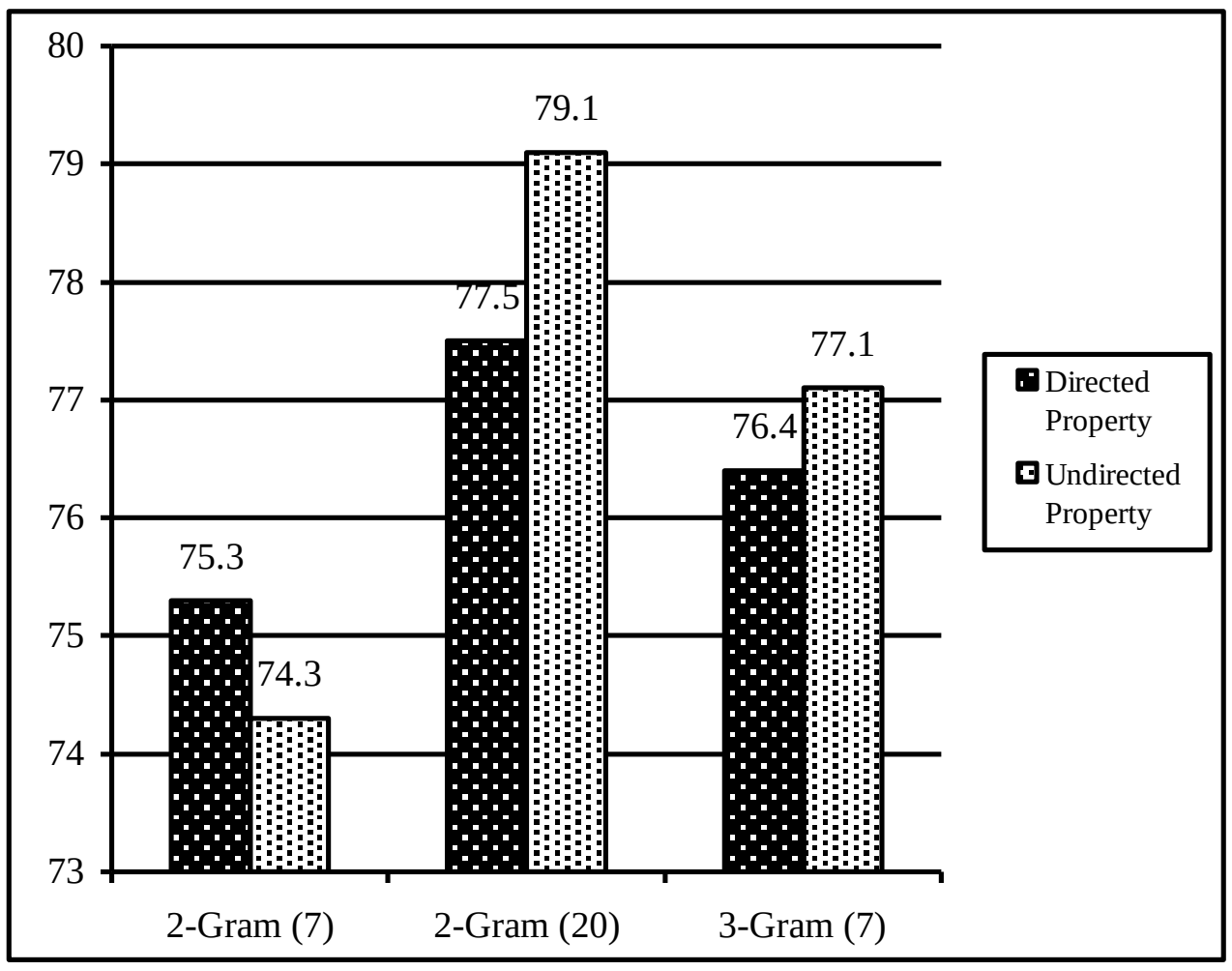

Fig. 4. Comparison of the F-Measure resulted by various feature encoding methods and properties.

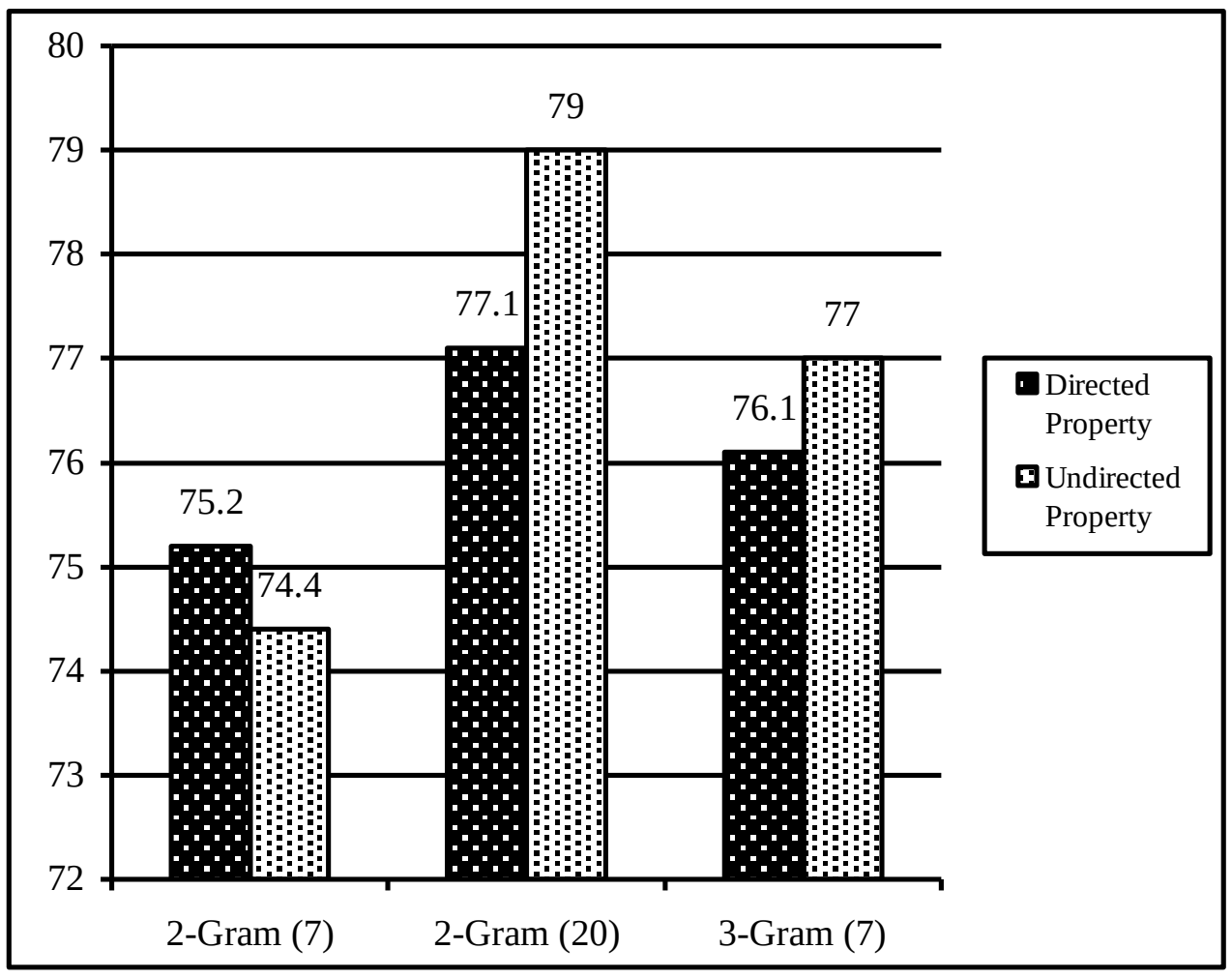

Fig. 5. Comparison of the Accuracy resulted by various feature encoding methods and properties.

As demonstrated in these figures, omitting the direction of amino acid increases the performance of encoding methods except 2-Gram(7). Since, in 2-Gram(7) and 3-Gram(7) encodings, when amino acids are categorized into seven classes, all members of each class are considered as identical and some useful information are missed. But, by comparing them with each other, 3-Gram(7) is a better encoding because it has the more information about amino acid compositions rather 2-Gram(7). Therefore, 2-Gram(7) cannot be a proper representative of proteins even with “undirected” property. Furthermore, 2-Gram(20) can perform better to demonstrate the sequence information. Thus, 2-Gram(20) encoding with “undirected” property has the best performance comparing to the other feature encodings. Finally, the performance of three types of proposed feature encoding methods are evaluated comparing to the Yu, Chou and Chang’s work [25] which significance vectors are constructed for sequence of proteins instead of the feature vectors; because, amino acid compositions which have critical role in protein-protein interactions are preserved in evolution. Fig.6 illustrates the superiority of 2-Gram(20) with “undirected” property by maximum *F-measure* of 79.1% and *Accuracy* of 79%.

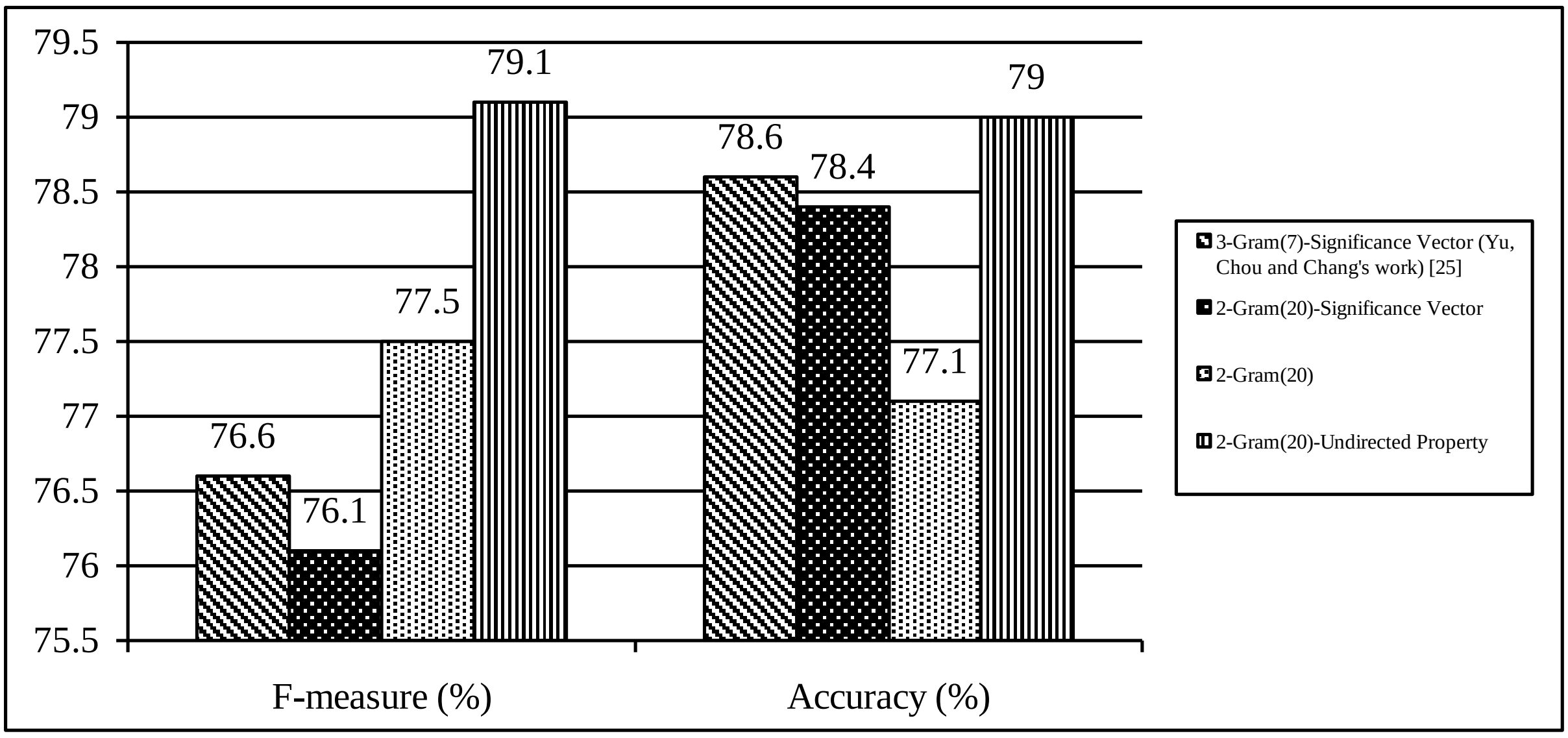

Fig. 6. Comparison of the F-measure and Accuracy resulted by various feature encoding methods with Yu, Chou and Chang’s work [25].